# Amplitude dependence of resonance frequency and its consequences for scanning probe microscopy

Omur E. Dagdeviren[1,*], Yoichi Miyahara[1], Aaron Mascaro[1], Tyler Enright[1], Peter Grütter[1]

[1] Department of Physics, McGill University, Montréal, Québec, Canada, H3A 2T8

*Corresponding author's email: omur.dagdeviren@mcgill.ca

## Abstract

With recent advances in scanning probe microscopy (SPM), it is now routine to determine the atomic structure of surfaces and molecules while quantifying the local tip-sample interaction potentials. Such *quantitative* experiments are based on the accurate measurement of the resonance frequency shift due to the tip-sample interaction. Here, we experimentally show that the resonance frequency of oscillating probes used for SPM experiments change systematically as a function of oscillation amplitude under typical operating conditions. This change in resonance frequency is not due to tip-sample interactions, but rather due to the cantilever strain or geometric effects and thus the resonance frequency being a function of the oscillation amplitude. Our numerical calculations demonstrate that the amplitude dependence of the resonance frequency is an additional yet overlooked systematic error source that can result nonnegligible errors in measured interaction potentials and forces. Our experimental results and complementary numerical calculations reveal that the frequency shift due to this amplitude dependence needs to be corrected even for experiments with active oscillation amplitude control to be able to quantify the tip-sample interaction potentials and forces with milli-electron volt and pico-Newton resolutions.

## Introduction

Resonant structures are widely used as accurate measurement devices in fields of science ranging from biological, chemical detection to gravitational waves or quantum mechanical systems [1-5]. Also, oscillating structures play a transducer role in atomic force microscopy (AFM) and related techniques [6, 7]. Dynamic AFM is an analytical surface characterization tool where a sharp probe tip is mounted to the end of an oscillating probe which serves as a frequency sensing element to measure surface properties with picometer, milli-electron volt, and pico-Newton resolution [6-8]. Among different dynamic AFM methodologies [6, 7, 9], frequency modulation AFM (FM-AFM) technique is the dominant high-resolution material characterization method under ultra-high vacuum conditions [7, 8, 10]. FM-AFM technique tracks the change of the resonance frequency $\Delta f$ of the cantilever under the influence of the attractive (or repulsive) surface forces while keeping the oscillation amplitude '*constant*' [8]. Even though the physical foundations of variations of the device resonance frequency was a long-lasting problem that was postulated in mid-1970s [11, 12], only recently it has been systematically demonstrated that the geometric and stress-induced deviations can modify the resonance frequency of micro-cantilever beams [13-16]. Here, we explore the amplitude dependence of resonance frequency for the most commonly used oscillating probes in dynamic SPM measurements: tuning forks, tuning forks in the qPlus configuration, and cantilever beams.

Our experimental results reveal that the resonance frequency of oscillating probes changes by a nonnegligible amount within practical operating conditions. The resonance frequency of tuning forks and tuning forks in the qPlus configuration decrease with increasing oscillation amplitude. The decrease in the resonance frequency implies that the amplitude dependence of the resonance frequency of tuning forks and tuning forks in the qPlus configuration is dominated by in-plane stress near the clamped end of the beam, i.e. surface stress effect [13-16]. In contrast, the resonance frequency of cantilever beams increases with the oscillation amplitude due to geometric effects, i.e. geometry change due to elastic deformation with the application of a load [14, 15]. We conducted numerical calculations to explore consequences of the amplitude dependence of the resonance frequency for high-resolution AFM measurements. Investigating the validity and the accuracy of potential energy (and/or force) spectroscopy experiments are ongoing research efforts due to their impact on the quantitative material characterization [17, 18]. For this reason, we concentrated on the force spectroscopy experiments which are performed to quantify the tip-sample interaction potential as a function of distance (and, by calculating its derivative, ultimately the force) with up to picometer precision laterally with respect to the tip's position relative to the location of surface atoms, often with the intent to explore the surface's chemical or electronic properties [19-21]. Our numerical calculations address that the amplitude dependence of the resonance frequency is an important yet overlooked systematic error source which can impede the measurement of the tip-sample interaction potential and force with milli-electron volt and pico-Newton resolutions. Therefore, the systematic error due to the amplitude dependence of the resonance frequency should be corrected for meaningful and accurate data acquisition even for well-posed interactions.

**Results and Discussions**

We experimentally investigated the variation of the resonance frequency of tuning forks, tuning forks in the qPlus configuration, and cantilevers as a function of oscillation amplitude (*see SI for details of experimental methods*). As Figure 1 shows, the resonance frequency of three different types of encapsulated tuning forks was measured with thermal noise spectra and frequency sweep experiments (*see Figure SI 1-3 for additional experimental results*). Oscillation amplitudes of tuning forks are calibrated with the principle of energy dissipation, details of which can be found elsewhere [22]. As highlighted in Figure 1, the resonance frequency of excited tuning forks decreases with respect to the thermal excitation measurements. The viscous effects of the surrounding medium can be excluded as tuning forks are encapsulated (i.e. in vacuum) [23]. Also, piezoelectric nonlinearities can be excluded as current-induced piezoelectric nonlinearities increase the resonance frequency of *z*-cut piezoelectric devices rather than decreasing the resonance frequency as we observed in our experiments [24, 25]. Both thermal noise density and frequency sweep experiments are conducted successively in a temperature controlled and quiet room in a thermally isolated chamber. With the other effects eliminated, the drop of the resonance frequency implies that the effect of in-plane surface stress is the governing factor for the amplitude dependence of the resonance frequency [13-16]. The first mode of tuning forks is along the vertical direction with respect to the sample surface, which induces the in-plane stress (i.e. the stress is in the same plane as the mechanical oscillation) near the clamp (*see SI for further discussion*). As proposed by former experimental and theoretical work, such in-plane stress results in a drop of the resonance frequency [13-16]. Also, we want to note that the multiplication of the relative frequency shift and the effective spring constant of the tuning forks ($k_{eff} = 2k$, where $k$ is the stiffness of a single prong) converge to $3,460 \pm 160$ (kg/s$^3$), which is another indication of the surface stress effect (also *see S.I. for further discussion*).

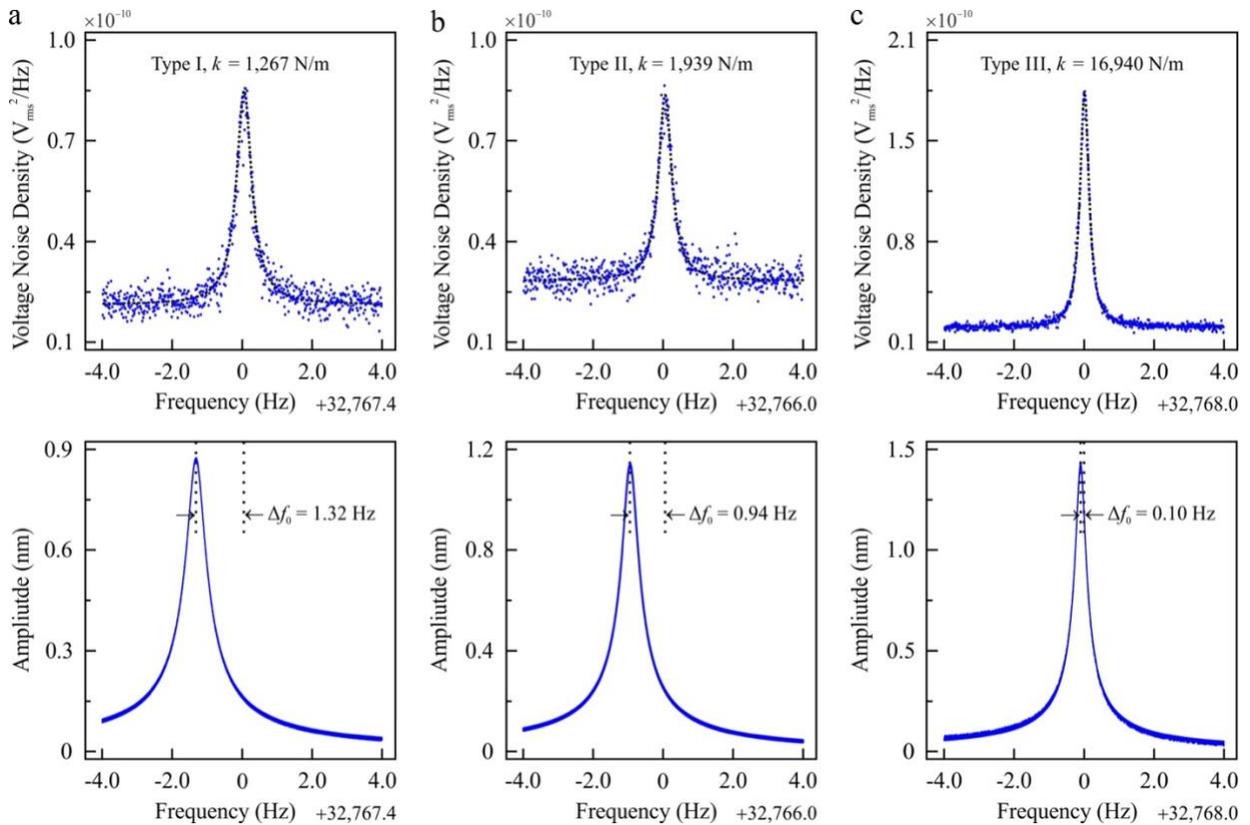

**Figure 1:** The measurement of the resonance frequency of encapsulated tuning forks with thermal noise density (first row. a-c) and frequency sweep experiments (second row, a-c). A Lorentzian curve is fitted to the thermal noise density and the resonance frequency is calculated from the fit. Thermal noise spectra presented in this figure are averaged 100 times. The oscillation amplitude of tuning forks is calibrated with the principle of energy dissipation [22].

Quartz tuning forks that have one free prong to which the tip is attached to the end while the fork's other prong is fixed to a holder ('qPlus' configuration) have gained popularity in recent years for high-resolution imaging [23]. Figure 2 discloses the decrease in resonance frequency with increasing oscillation amplitude of a qPlus sensor. We conducted successive frequency sweep experiments in which we increased and decreased the frequency to investigate the potential contribution of duffing nonlinearity to the change in resonance frequency [26]. We conducted experiments with increasing and decreasing oscillation amplitudes. The identical same resonance curves were found, establishing our system as a harmonic oscillator. The results in Figure 2 (*also Figure SI 1-3 for additional experimental data*) show that the resonance frequency is a function of amplitude even in the small oscillation amplitude range (Ångströms to nanometers). As Figure 2 shows, the variation of the resonance frequency is more emphasized for tuning forks in the qPlus configuration compared to encapsulated tuning forks (*see Figure 1 and also Figure SI 1-3*). The increased effect of the oscillation amplitude on the resonance frequency can be linked to the sensor assembly, which can alter the stress concentration near the clamp [13].

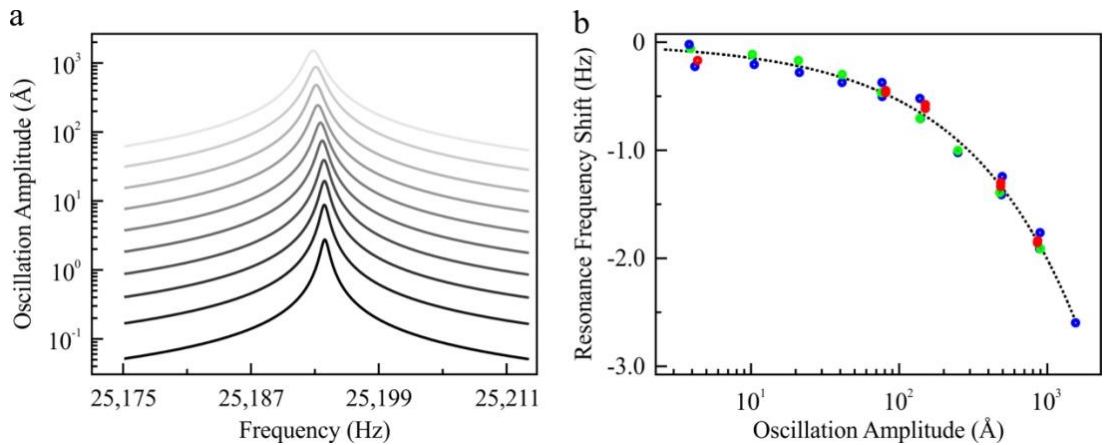

**Figure 2:** The measurement of the resonance frequency of a tuning fork (type-II) in the qPlus configuration. Frequency sweep experiments (a) are conducted to determine the resonance frequency. The oscillation amplitude of the qPlus is calibrated with the principle of energy dissipation [22]. We fit a curve to experimental data (black, dashed line), $\Delta f = -0.03548 \times A^{0.584}$, where $\Delta f$ is the resonance frequency shift in Hz and $A$ is the oscillation amplitude in Ångströms. To eliminate the duffing nonlinearity and other sources of nonlinearities (e.g. temperature change due to oscillation), we conducted successive experiments under different excitation conditions. We increased the resonance frequency during the sweep experiments and the oscillation amplitude for successive sweeps (blue circles). Also, we decreased the frequency during resonance sweep experiments (red circles) and we decreased the oscillation amplitude for each consecutive experiment (green circles). To eliminate the effect of viscous damping and temperature variations, experiments presented in this figure are conducted at liquid nitrogen temperature in vacuum conditions.

For completeness, we also performed measurements of the amplitude dependence of the resonance frequency with conventional micro-fabricated silicon cantilevers which are widely used for scanning probe microscopy experiments [6, 7]. We compared thermal noise spectrum and frequency sweep experiments for two different types of cantilevers (cantilever-I OPUS 4XC-NN-A, cantilever-II OPUS 4XC-NN-B) in vacuum. As Table 1 (*also Figure SI 4 for additional experimental data*) summarizes, the resonance frequency of cantilevers increases with increasing oscillation amplitude in contrast to encapsulated tuning forks and tuning forks in the qPlus configuration, where the resonance frequency decreases with the amplitude. Our experiments with cantilever beams are in line with former experimental and theoretical work which proposed that resonance frequency of cantilever beams increases due to geometric effects [14, 15].

|  | $f_0$ with Thermal Noise (Hz) | $f_0$ with Frequency Sweep (Hz) |
|---|---|---|
| **Cantilever-I** | 18156.8 ± 0.4 | 18163.4 ± 0.2 |
| **Cantilever-II** | 84325.9 ± 1.5 | 84329.0 ± 1.5 |

**Table 1:** The measurement of the resonance frequency of two different types of micro-cantilevers with thermal noise density and frequency sweep experiments. For both cantilevers, the resonance frequency increases with the oscillation amplitude. Experimental data are the average of 5 independent measurements of 50 spectra averaged together. The resonance frequency is determined by fitting a Lorentzian function to the thermal noise spectrum. The amplitude at the resonance frequency is 8.2 nm for cantilever-I and 7.5 nm for cantilever-II. The vacuum system is kept in a quiet room during the experiments to eliminate the effect of acoustic noise [27].

In the following, we investigate what the consequences are in the field of non-contact AFM (NC-AFM) for the amplitude dependent resonant frequency. In NC-AFM, measurements of the resonance frequency as a function of tip-sample separation are often used to reconstruct the sample's potential energy landscape. This is known potential energy (and/or force) spectroscopy and used to understand reaction pathways, diffusion, as well as validate *ab-initio* modeling. To demonstrate the effect of the amplitude dependent frequency, we conducted numerical calculations to determine the consequences on the potential energy and force spectroscopy experiments. Even though the amplitude is intended to be '*kept constant*' via feedback mechanism during the frequency modulation-based force spectroscopy experiments, a non-zero amplitude error is always present [8]. To understand the effect of the amplitude error, we fitted an empirical curve to the experimental data presented in Figure 2b. We use Equation 1 to calculate the resonance frequency shift due to the amplitude error around the amplitude setpoint:

$$\Delta f_{error} = \Delta f_{fit}(A_{set}) - \Delta f_{fit}(A_{set} + A_{error}) \qquad (1)$$

where $\Delta f_{error}$ is the frequency shift due to amplitude dependence of resonance frequency, $A_{set}$ is the amplitude setpoint, and $A_{error}$ is the amplitude error using the fitted curve ($\Delta f_{fit}$) to the experimental data in Figure 2b. We selected an amplitude error up to 10 picometers and assumed a constant $A_{error}$ as a function of tip-sample distance. Note that in reality, however, $A_{error}$ can increase more significantly in the proximity of the surface and can have a larger error corrugation across the scan area due to non-linear tip-sample interaction and due to the effect of dissipative forces [9, 28]. For this reason, the assumption of a constant $A_{error}$ underestimates the amplitude error with respect to experimental conditions.

We calculated the resonance frequency shift due to a tip-sample interaction potential as a function of $A_{set}$ and then reconstructed the tip-sample interaction potential using well-established mathematical procedures with and without considering the effect of $A_{error}$ on the frequency shift (*see S.I. for details of numerical calculations*) [29]. As Figure 3 summarizes, we investigated the error in the reconstructed potential energy and force as a function of amplitude error ($A_{error}$), amplitude setpoint ($A_{set}$), and tip-sample distance. As Figure 3 a and b reveals, unless the amplitude error is minimized or corrected, the amplitude dependence of the resonance frequency, $\Delta f_{error}$, does not allow measurements with milli-electron volt and pico-Newton accuracy. We emphasize that the systematic error due to the amplitude dependence of the resonance frequency is a significant, yet overlooked source of experimental error given the fact that the corrugation of the tip-sample interaction potential of high-resolution spectroscopy experiments is on the order of a few tens of milli-electron volts and force corrugation around hundred pico-Newtons [19, 21, 30]. As highlighted by Figure 3, the error of the reconstructed tip-sample interaction potential and force peaks when $A_{set}$ is comparable to $A_{error}$. In addition to this trivial trend, the effect of amplitude error increases when $A_{set}$ is on the order of nanometers as the amplitude-dependent change in resonance frequency, $\Delta f_{error}$, inflates as presented in Figure 2 (*also see Figure SI 1-3*).

Our results highlight that the amplitude dependence of the resonance frequency is a systematic error source due to the intrinsic properties of the oscillating probe used, independent of the nature of tip-sample interaction potential. For this reason, the error induced due to the amplitude dependence of the resonance frequency will persist even for a well-posed interaction [17, 18]. As a consequence, most experimental force spectroscopy data derived from frequency shift measurements presented in the literature is expected to have systematic errors that can be corrected for. As equation 2 demonstrates, by measuring the oscillation amplitude ($A_{osc}$) dependence of the resonance frequency for the sensor in use, e.g. Figure 2b, and the oscillation amplitude error, the contribution of amplitude error on quantitative spectroscopy experiments can be eliminated.

$$\Delta f_{tip-sample} = \Delta f_{measured} - \Delta f_{error} \qquad (2)$$

Our numerical calculations show that unless the resonance frequency shift due to amplitude dependent resonance frequency error is corrected, the oscillation amplitude error has to be kept less than one picometer to achieve milli-electron volt and piconewton resolution. Controlling the oscillation amplitude with sub-picometer error corrugation across the scan area is not straightforward at all due to hardware related experimental limitations [7, 9].

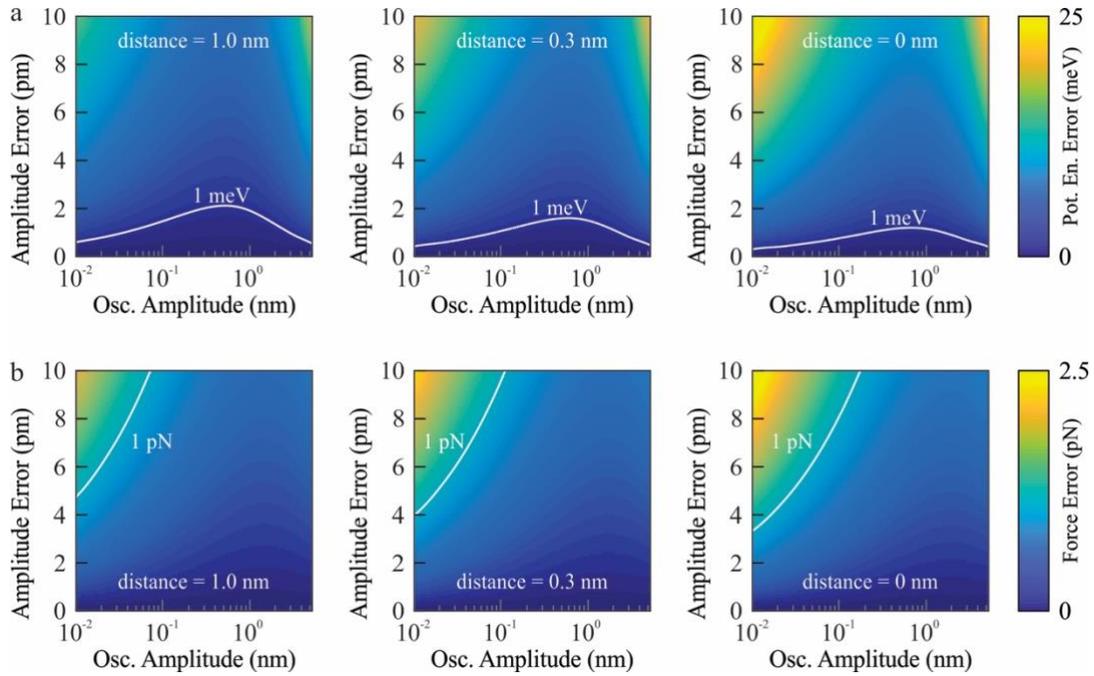

**Figure 3:** The effect of amplitude dependence of the resonance frequency on recovered tip-sample interaction potential (a) and force (b) of tip-sample distance and amplitude error. (a) The error of recovered tip-sample interaction potential increases with decreasing tip-sample separation. To keep the error less than one milli-electron volt and one pico-Newton (white contour in (a) and (b)), the amplitude error should be less than one picometer. If the amplitude error is not constraint to sub-picometer range, the frequency shift due to the amplitude dependence of the resonance frequency should be corrected. By this way, the intrinsic contribution of frequency shift due to tip-sample interaction can be calculated and error in the reconstructed tip-sample interaction potential and force can be eliminated. The reference point for the distance, i.e. distance = 0, is defined as the potential energy minimum of the model tip-sample interaction (*see S.I. for details*).

We want to emphasize that even if the amplitude dependence of the resonance frequency has a different physical origin other than the surface stress or geometric effects described, the consequences on the spectroscopy experiments would still be valid. Also, possible implications of the amplitude dependent resonance frequency are not limited to force spectroscopy experiments. For instance, multimodal atomic force microscopy in which the relative amplitudes of different harmonics can, in potential, change due to the amplitude dependence of resonance frequency. With the change of the modal shape, the amplitude dependence of the resonance frequency of higher harmonics can be different than each other. This may further complicate the data interpretation and the correction. Also, the optical drive or the thermal drive of oscillating probes may enhance non-linearities. This can promote additional non-linearity mechanisms due to temperature fluctuations. Moreover, Kelvin probe force microscopy and pump-probe atomic force microscopy measurements require a precise measurement of the resonance frequency. For this reason, the amplitude dependence of the resonance frequency can hamper quantitative Kelvin Probe force microscopy and pump-probe AFM measurements [31].

## Summary

We experimentally demonstrated that the resonance frequency of the most commonly used oscillating probes for dynamic atomic force microscopy experiments changes significantly as a function of oscillation amplitude. Amplitude errors need to be maintained below one picometer to keep resultant systematic energy and force errors below milli-electron volt and pico-Newton, respectively. Our results demonstrate that the amplitude dependence of the resonance frequency is nonnegligible yet overlooked systematic error source for the quantitative measurement of tip-sample interaction potentials and forces if milli-electron volt and pico-Newton accuracy are claimed. Finally, we want to note that the amplitude dependence of the resonance frequency has possible consequences for other scanning probe microscopy methodologies.

## Acknowledgements

Financial support from The Natural Sciences and Engineering Research Council of Canada and Le Fonds de Recherche du Québec - Nature et Technologies are gratefully acknowledged.

## References

1. *Springer Handbook of Nano-technology*. 3rd Edition ed, ed. B. Bhushan. 2010, Springer Berlin Heidelberg.
2. J.L. Arlett, E.B. Myers, and M.L. Roukes, Comparative advantages of mechanical biosensors. Nature Nanotechnology **6**, 203 (2011).
3. B. Anja, D. Søren, K. Stephan Sylvest, S. Silvan, and T. Maria, Cantilever-like micromechanical sensors. Reports on Progress in Physics **74**, 036101 (2011).
4. B. P. Abbott et al. (LIGO Scientific Collaboration and Virgo Collaboration), Observation of Gravitational Waves from a Binary Black Hole Merger. Phys. Rev. Lett. **116**, 061102 (2016).
5. K.C. Schwab and M.L. Roukes, Putting Mechanics into Quantum Mechanics. Physics Today **58**, 36 (2005).
6. R. Garcia, *Amplitude Modulation Atomic Force Microscopy*. 2010, Singapore: Wiley-VCH.
7. F.J. Giessibl, Advances in atomic force microscopy. Reviews of Modern Physics **75**, 949-983 (2003).
8. T.R. Albrecht, P. Grutter, D. Horne, and D. Rugar, Frequency modulation detection using high-Q cantilevers for enhanced force microscope sensitivity. Journal of Applied Physics **69**, (1991).
9. O.E. Dagdeviren, J. Götzen, H. Hölscher, E.I. Altman, and U.D. Schwarz, Robust high-resolution imaging and quantitative force measurement with tuned-oscillator atomic force microscopy. Nanotechnology **27**, 065703 (2016).
10. F.J. Giessibl, AFM's path to atomic resolution. Materials Today **8**, 32-41 (2005).
11. J. Lagowski, H.C. Gatos, and E.S. Sproles, Surface stress and the normal mode of vibration of thin crystals :GaAs. Applied Physics Letters **26**, 493-495 (1975).
12. M.E. Gurtin, X. Markenscoff, and R.N. Thurston, Effect of surface stress on the natural frequency of thin crystals. Applied Physics Letters **29**, 529-530 (1976).

13. M.J. Lachut and J.E. Sader, Effect of Surface Stress on the Stiffness of Cantilever Plates. Physical Review Letters **99**, 206102 (2007).

14. R.B. Karabalin, L.G. Villanueva, M.H. Matheny, J.E. Sader, and M.L. Roukes, Stress-Induced Variations in the Stiffness of Micro- and Nanocantilever Beams. Physical Review Letters **108**, 236101 (2012).

15. M.J. Lachut and J.E. Sader, Effect of surface stress on the stiffness of thin elastic plates and beams. Physical Review B **85**, 085440 (2012).

16. J.J. Ruz, V. Pini, O. Malvar, P.M. Kosaka, M. Calleja, and J. Tamayo, Effect of surface stress induced curvature on the eigenfrequencies of microcantilever plates. AIP Advances **8**, 105213 (2018).

17. J.E. Sader, B.D. Hughes, F. Huber, and F.J. Giessibl, *Interatomic force laws that corrupt their own measurement*, in *arXiv:1709.07571*. 2017.

18. O.E. Dagdeviren, C. Zhou, E.I. Altman, and U.D. Schwarz, Quantifying Tip-Sample Interactions in Vacuum Using Cantilever-Based Sensors: An Analysis. Physical Review Applied **9**, 044040 (2018).

19. L. Gross, F. Mohn, N. Moll, P. Liljeroth, and G. Meyer, The Chemical Structure of a Molecule Resolved by Atomic Force Microscopy. Science **325**, 1110-1114 (2009).

20. L. Gross, F. Mohn, P. Liljeroth, J. Repp, F.J. Giessibl, and G. Meyer, Measuring the Charge State of an Adatom with Noncontact Atomic Force Microscopy. Science **324**, 1428-1431 (2009).

21. B.J. Albers, T.C. Schwendemann, M.Z. Baykara, N. Pilet, M. Liebmann, E.I. Altman, and U.D. Schwarz, Three-dimensional imaging of short-range chemical forces with picometre resolution. Nat Nano **4**, 307-310 (2009).

22. O.E. Dagdeviren, Y. Miyahara, A. Mascaro, and P. Grutter, Calibration of the oscillation amplitude of electrically excited scanning probe microscopy sensors. arXiv:1809.01584 (2018).

23. F.J. Giessibl, High-speed force sensor for force microscopy and profilometry utilizing a quartz tuning fork. Applied Physics Letters **73**, 3956-3958 (1998).

24. J.J. Gagnepain and R. Besson, *5 - Nonlinear Effects in Piezoelectric Quartz Crystals*, in *Physical Acoustics*, W.P. Mason and R.N. Thurston, Editors. 1975, Academic Press. p. 245-288.

25. R.E. Newnham, The Amplitude-Frequency Effect in Quartz Resonators. Ferroelectrics **306**, 211-220 (2004).

26. A.A. Vives, *Piezoelectric Transducers and Applications*, ed. A.A. Vives. 2008: Springer, Berlin, Heidelberg.

27. A. Mascaro, Y. Miyahara, O.E. Dagdeviren, and P. Grütter, Eliminating the effect of acoustic noise on cantilever spring constant calibration. Applied Physics Letters **113**, 233105 (2018).

28. A. Labuda, Y. Miyahara, L. Cockins, and P.H. Grütter, Decoupling conservative and dissipative forces in frequency modulation atomic force microscopy. Physical Review B **84**, 125433 (2011).

29. J.E. Sader and S.P. Jarvis, Accurate formulas for interaction force and energy in frequency modulation force spectroscopy. Applied Physics Letters **84**, 1801-1803 (2004).

30. A.J. Weymouth, T. Hofmann, and F.J. Giessibl, Quantifying Molecular Stiffness and Interaction with Lateral Force Microscopy. Science **343**, 1120 (2014).

31. Z. Schumacher, A. Spielhofer, Y. Miyahara, and P. Grutter, The limit of time resolution in frequency modulation atomic force microscopy by a pump-probe approach. Applied Physics Letters **110**, 053111 (2017).

# Supplemental Material:

## Amplitude dependence of resonance frequency and its consequences for scanning probe microscopy

Omur E. Dagdeviren[1,*], Yoichi Miyahara[1], Aaron Mascaro[1], Tyler Enright[1], Peter Grütter[1]

[1] Department of Physics, McGill University, Montréal, Québec, Canada, H3A 2T8

*Corresponding author's email: omur.dagdeviren@mcgill.ca

## I. Experimental methods and additional experimental results

We conducted thermal noise spectra experiments and frequency sweep experiments to measure the resonance frequency of probes. The thermal noise spectra were obtained by fast Fourier transform of the detection signal. The driven frequency spectra were measured by a lock-in amplifier with a sinusoidal drive signal with constant amplitude while its frequency is swept. The frequency resolution of the thermal noise spectra is 4 milli-Hz. The frequency resolution of the driven frequency spectra is 0.4 milli-Hz for encapsulated tuning forks, 10 milli-Hz for tuning forks in the qPlus configuration, and 0.1 Hz for cantilevers. The resonance frequency is determined by fitting a Lorentzian curve to the experimental frequency spectra around the peak. The oscillation amplitude of tuning forks and tuning forks in the qPlus configuration are measured with the principle of energy dissipation details of which can be found elsewhere [1]. All experiments are conducted in a quiet room to avoid the contribution of acoustic noise [2].

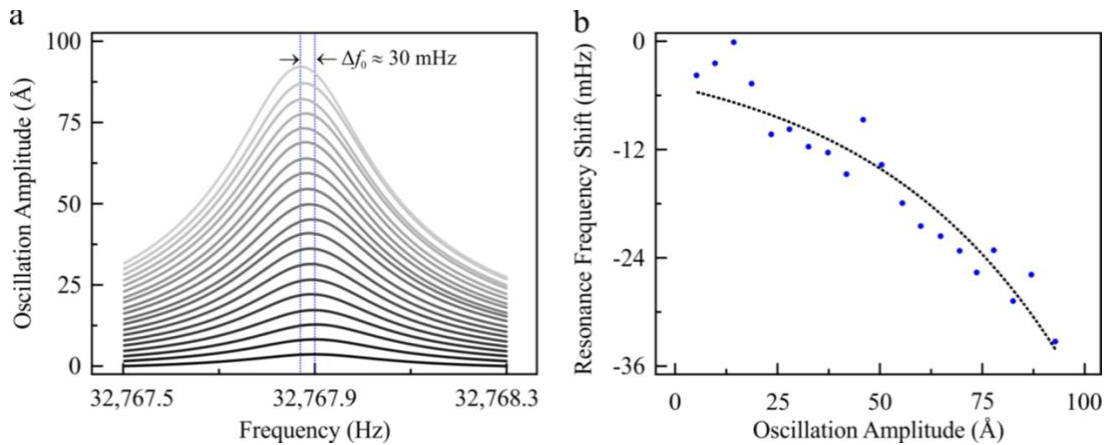

**Figure SI 1:** The measurement of the resonance frequency of an encapsulated tuning fork as a function of oscillation amplitude. (a) Frequency sweep experiments are conducted to determine the resonance frequency. The oscillation amplitude of the tuning fork (type-III) is calibrated with the principle of energy dissipation [1]. (b) The resonance frequency decreases with increasing oscillation amplitude. To eliminate the effect of temperature variations, experiments presented in this figure are conducted at constant temperature in a thermally isolated chamber.

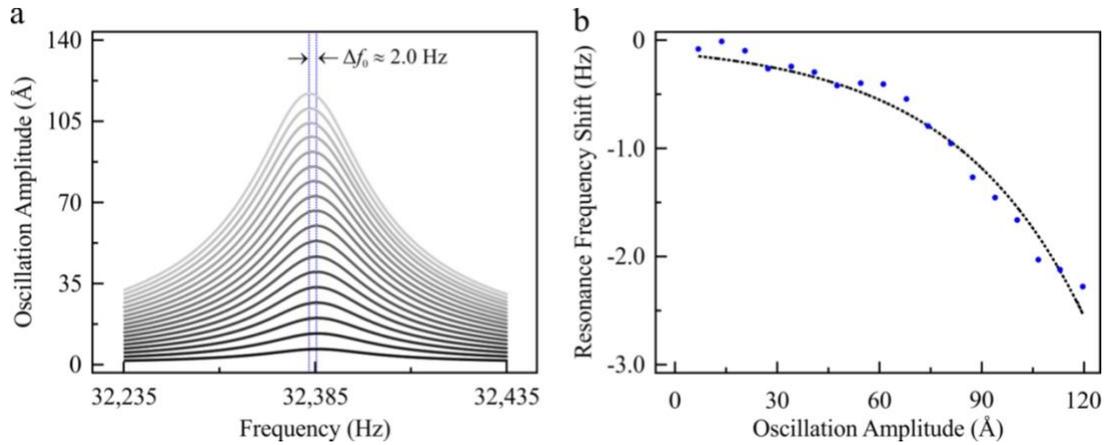

**Figure SI 2:** The measurement of the resonance frequency of a tuning fork in qPlus configuration. (a) Frequency sweep experiments are conducted to determine the resonance frequency. The oscillation amplitude of the qPlus sensor (based on tuning fork type-III) is calibrated with the principle of energy dissipation [1]. (b) The resonance frequency changes up to 2 Hz when the oscillation amplitude is changed from 8 Å to 120 Å. The enhanced amplitude dependence is expected to be associated with the increased stress concentration at the clamped end of the prong due to assembly process. To eliminate the effect of temperature variations, experiments presented in this figure are conducted at constant temperature in a thermally isolated chamber.

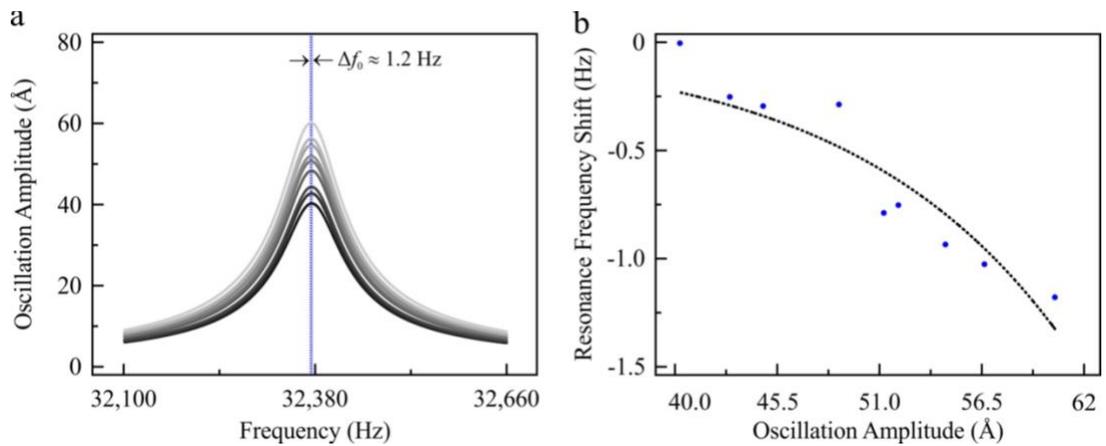

**Figure SI 3:** The measurement of the resonance frequency of a tuning fork in qPlus configuration. (a) Frequency sweep experiments are conducted to determine the resonance frequency. The oscillation amplitude of the qPlus sensor (based on tuning fork type-II) is calibrated with the principle of energy dissipation [1]. (b) The resonance frequency changes up to 1.2 Hz when the oscillation amplitude is tuned up ~20 Å. The enhanced amplitude dependence is expected to be associated with the increased stress concentration at the clamped end of the prong due to assembly process [3, 4]. To eliminate the effect of temperature variations, experiments presented in this figure are conducted at room temperature in a thermally isolated chamber.

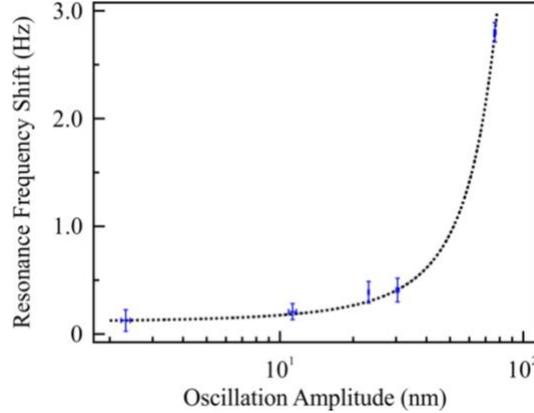

**Figure SI 4:** The measurement of the resonance frequency of cantilever-I as a function of oscillation amplitude. The resonance frequency of the cantilever increases with the oscillation amplitude, which implies that geometric effects dominate [5]. Experiments are conducted under ambient conditions in a temperature-controlled room. To eliminate the effect of acoustic noise, experiments are performed in a quiet room [2].

## II.  In-plane stress near the tuning-fork base

We briefly summarize the finite element method (FEM) approach to calculate the modal shape of the oscillation and the stress distribution of oscillating tuning fork-based sensors while further details for FEM calculations can be found elsewhere [6-10]. All calculations were performed using COMSOL Multiphysics 4.4 structural mechanics software package (COMSOL Multiphysics, GmbH, Berlin-Germany).

The physical system is modeled when FEM calculations are conducted. Modeling the tuning fork for FEM calculations requires measuring the dimensions. We employed a calibrated optical microscope to obtain the dimensions of the tuning forks (Table SI 1). It is important to reflect the tuning fork's geometry accurately in regions where stress concentrations are expected, e.g. the region between the prongs and where prongs are connected to the base of the tuning fork [9, 10]. We did not include the gold coating, which has an average thickness of 200 Å, nor the notches at the tuning fork base to decrease the cost of computation and modeling. Neglecting these features in our FEM model has been justified as they are mainly important for electrical properties of tuning fork while having no substantial influence on the mechanical properties [8, 9, 11].

|                  | Width      | Length of the Prong | Thickness  |
|------------------|------------|---------------------|------------|
| **Tuning Fork-I**   | 234 (μm)   | 2471 (μm)           | 90 (μm)    |
| **Tuning Fork-II**  | 234 (μm)   | 2426 (μm)           | 131 (μm)   |
| **Tuning Fork-III** | 600 (μm)   | 3600 (μm)           | 250 (μm)   |

**Table SI 1:** Dimensions of tuning forks are measured with a calibrated light microscope.

In addition to measuring geometric dimensions, assigning relevant material properties such as Young's modulus, Poisson's ratio, mass density, and damping coefficient are required. We used the material properties for quartz from the materials library of the FEM software with the mass density of a common epoxy glue (Epoxy Technology's EPO-TEK H72) as derived from the information provided in the manufacturer's data sheet and the damping coefficient of quartz and epoxy taken from Ref [12, 13] (Table SI 2). Finally, we assumed that Macor does not contribute to the damping to speed up the calculations; doing so is justified due to the much smaller damping coefficient of Macor compared to quartz and, in particular, the epoxy, damping due to Macor has virtually no influence on the simulation results [9, 10].

|  | Young's modulus (Gpa) | Poisson's ratio | Mass density (kg m$^{-3}$) | Damping Coefficient |
|---|---|---|---|---|
| **Quartz** | 82 | 0.17 | 2648 | $2 \times 10^{-4}$ |
| **Epoxy** | 7 | 0.35 | 1600 | $5 \times 10^{-3}$ |
| **Macor** | 300 | 0.222 | 3900 | - |

**Table SI 2:** Material properties used for finite element method (FEM) calculations.

As Figure SI 5a summarizes, to determine the modal shape and the surface stress distribution due to the mechanical oscillation, boundary conditions are defined at the base of Macor holder. Figure SI 5b shows that we increased the mesh density at the material interfaces and at regions where stress concentrations are expected. Eigenfrequency analysis of the assembly system reveals that the resonance frequency of the tuning fork assembly is 32,171 Hz and tuning fork oscillates along the vertical direction. As Figure SI 5c discloses that in-plane stress distribution is evident along the oscillation direction (with brighter colors presenting higher stress). It is clearly seen that stress is highest near the fixed end of the cantilevered free prong along the oscillation direction (in-plane) and can, in potential, induce variations of the resonance frequency [3-5, 14].

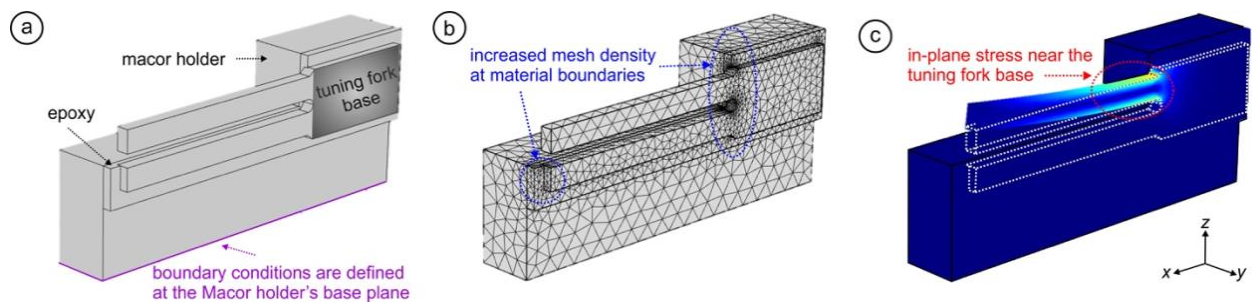

**Figure SI 5:** The assembled sensor setup and the modelling procedure to determine the modal shape and the surface stress distribution of tuning fork-based sensors. (a) Three-dimensional presentation of the sensor assembly including the tuning fork, Macor holder, and epoxy glue used to fix the tuning fork to the base holder. (b) Enhanced mesh density is evident at the material boundaries and at the areas of stress concentrations. (c) The modal shape of the oscillation is calculated with eigenfrequency analysis. The first operational mode of the sensor assembly has a resonance frequency of 32,171 Hz and the oscillation is along *z*-direction. With brighter colors representing higher stress along

the oscillation direction (in-plane), it is evident that the stress is the highest near the fixed end of the cantilevered free prong.

### III. Multiplication of the spring constant with the relative resonance frequency shift

It has been previously shown that the relative frequency shift due to surface stress of beams can be expressed [3-5]:

$$\frac{\Delta_w}{w_0} = -0.042 \frac{v(1-v)\sigma_s^T}{k_{ref}}, \tag{1}$$

In equation 1, $\Delta_w$ is the relative frequency shift, $w_0$ is the resonant frequency in the absence of the surface stress, $v$ is the Poisson's ratio, $\sigma_s^T$ is the total surface stress. Equation 1 implies that the multiplication of $\Delta_w$ with the reference stiffness of the cantilever beam, $k_{ref}$ remains constant for the same range of total surface stress.

To investigate the convergence of the surface stress when the oscillation amplitude is increased from thermal limit to nanometer range, we calculated the stiffness of the tuning forks we used in our experiments. Experimental and theoretical approaches are available to calibrate the spring constant ($k$) of the probes used for scanning probe microscopy (SPM) experiments (see Ref [15-18] for detailed reviews). We used FEM-based technique to calculate the spring constant of tuning forks (*see section II of the SI for details*). As Figure SI 6a summarizes, while keeping one of the prongs and the base of the tuning fork rigid, we applied force along *z*-direction to the end of the free prong. We swept the force from 1 μN to 100 μN with 1 μN steps and measured the displacement, Δ*z*. The next step is fitting a first-order polynomial to find the spring constant of the tuning fork by using Hooke's law.

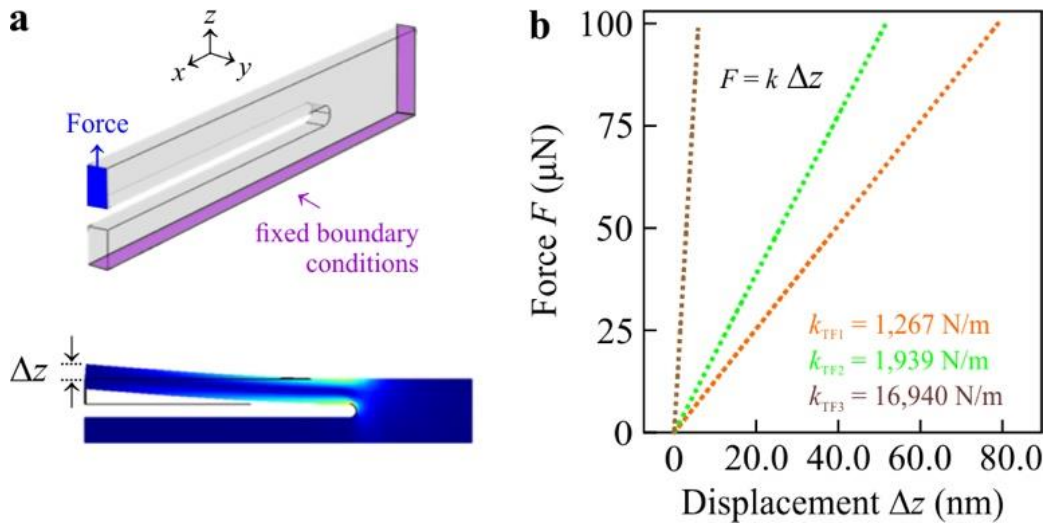

**Figure SI 6:** Calibration of spring constant with finite element methods. (a) One of the prongs and the base of the tuning fork are fixed, while force is applied to the end of the free prong along the *z*-direction. The exerted force

deforms the prong and the displacement at the end of the prong, Δz, is measured. (b) The slope of the force versus displacement curve is equal to the spring constant of the tuning fork, k.

Summarized in Figure SI 6b, we find consistent spring constant values with earlier experimental and computational results for similar tuning forks [6-8, 19, 20]. As outlined by different groups, calibration of spring constant with FEM techniques deviates up to 5% with respect to dynamic experimental results [6, 9, 10, 21].

As Figure 1 in the main text shows, tuning forks are excited with very similar oscillation amplitudes, i.e. all in nanometer range. For this reason, the total surface stress is assumed to be in the same range. The multiplication of the relative resonance frequency shift (Figure 1 in the main text) with the equivalent spring constant of tuning forks ($k_{eff} = 2k$, where $k$ is the stiffness of a single prong) calculated with finite element method gives a constant product of 3,460 ± 160 (kg/s$^3$). Such a convergence implies that the in-plane surface stress is the governing physical phenomena that leads the amplitude dependent resonance frequency. In addition, we want to underline that even if the amplitude dependence of the resonance frequency has a different physical origin other than the surface stress effect (for tuning forks) or geometric effect (for cantilevers), consequences on the spectroscopy experiments will always exist.

IV. Computational methods employed in the numerical analysis of resonance shifts

Following a commonly used approach for dynamic atomic force microscopy simulations, we solved the equation of motion of a damped harmonic oscillator with external excitation and non-linear tip-sample interaction force [22-26]:

$$m\ddot{z}(t) + \frac{2\pi f_0 m}{Q}\dot{z}(t) + c_z[z(t) - d] = a_d c_z \cos(2\pi f_d t) + F_{ts}[z(t), \dot{z}(t)] \ , \qquad (2)$$

where $z(t)$ is the position of the tip as a function of time $t$ (with $z = d$ denoting the distance of the tip relative to the sample when the cantilever is undeflected); $m$, $f_0$, $Q$, and $c_z$ are the effective mass, the first eigenfrequency, the quality factor, and the spring constant of the oscillator, respectively. In equation 2, the terms on the left reflect the standard terms for a damped harmonic oscillator, while the first term on the right represents the external excitation of the oscillator with excitation amplitude $a_d$ and excitation frequency $f_d$. The second term on the right side finally symbolizes the non-linear tip sample interaction force $F_{ts}$, which may depend both on the tip's time-dependent position $z$ as well as its instantaneous velocity $\dot{z}$. Neglecting a possible velocity dependence, we chose $F_{ts}$ in agreement with previous literature [24-26] as a combination of a van der Waals-type sphere-over-flat interaction [27] for the attractive regime ($z \geq z_0$) and a contact force ($z < z_0$) that follows Maugis' approximation to the Derjaguin-Muller-Toporov model (DMT-M) [28-30], which is often referred to as Hertz-plus-offset model [31]:

$$F_{ts}(z) = \begin{cases} -\dfrac{A_H R}{6z^2} & \text{for } z \geq z_0 \\ \dfrac{4}{3} E^* \sqrt{R} (z_0 - z)^{3/2} - \dfrac{A_H R}{6z_0^2} & \text{for } z < z_0 \end{cases} \qquad (3)$$

where $A_H$ = 0.2 aJ is the Hamaker constant, $R$ = 10 nm the radius of the tip's apex, $z_0$ = 0.3 nm the distance at which the contact is established, and $E^* = ((1-\nu_t^2)/E_t + (1-\nu_s^2)/E_s)^{-1}$ the combined elastic modulus of the tip and sample (with $E_t$ = 130 GPa as the Young's modulus of the tip, $E_s$ = 1 GPa as the Young's modulus of the sample, and $\nu_t = \nu_s = 0.3$ as the Poisson ratios of tip and sample, respectively). To describe the oscillator, we chose $c_z$ = 2000 N/m, $f_d = f_0$ = 22,000 Hz, $Q$ = 10,000; these values reflect typical parameters for a tuning fork glued on a holder in qPlus configuration, which represents the currently most common oscillator choice for high-resolution, vacuum-based atomic force microscopy.

Equation 2 was then solved by employing a previously derived analytical solution for the tip-sample motion, which is, however, defined for conservative tip-sample interactions only [25]. Finally, Equation 4 details the numerical integration method we used for reconstructing the tip-sample interaction potential $U_{ts}$ from data obtained with FM-type force spectroscopy introduced by Sader and Jarvis, which represents the most widely used reconstruction protocol for this case [32]. It results in the following equation:

$$U_{ts}(D) = 2c_z \int_D^\infty \frac{f_0 - f_{res}}{f_0} \left[ (z-D) + \sqrt{\frac{A}{16\pi}} \sqrt{z-D} + \frac{A^{3/2}}{\sqrt{2(z-D)}} \right] dz, \qquad (4)$$

Note that $U_{ts}$ is given as a function of nearest tip-sample distance $D$, which distinguishes itself from the distance $d$ the tip has to the surface when the cantilever is undeflected by $D = d - A$, where $f_{res}(D)$ represents the cantilever's distance-dependent resonance frequency (i.e., $\Delta f = f_0 - f_{res}$). With the knowledge of $U_{ts}(D)$, the tip-sample force $F_{ts}$ as a function of $D$ can easily be recovered for both cases by calculating its negative gradient ($F_{ts}(D) = -\partial U_{ts}/\partial D$).

**References for Supplemental Material**

1. O.E. Dagdeviren, Y. Miyahara, A. Mascaro, and P. Grutter, Calibration of the oscillation amplitude of electrically excited scanning probe microscopy sensors. arXiv:1809.01584 (2018).
2. A. Mascaro, Y. Miyahara, O.E. Dagdeviren, and P. Grütter, Eliminating the effect of acoustic noise on cantilever spring constant calibration. Applied Physics Letters **113**, 233105 (2018).
3. M.J. Lachut and J.E. Sader, Effect of Surface Stress on the Stiffness of Cantilever Plates. Physical Review Letters **99**, 206102 (2007).

4. M.J. Lachut and J.E. Sader, Effect of surface stress on the stiffness of thin elastic plates and beams. Physical Review B **85**, 085440 (2012).

5. R.B. Karabalin, L.G. Villanueva, M.H. Matheny, J.E. Sader, and M.L. Roukes, Stress-Induced Variations in the Stiffness of Micro- and Nanocantilever Beams. Physical Review Letters **108**, 236101 (2012).

6. S. Georg Hermann, H. Markus, and R. Hans-Peter, Recipes for cantilever parameter determination in dynamic force spectroscopy: spring constant and amplitude. Nanotechnology **18**, 255503 (2007).

7. D. van Vörden, M. Lange, M. Schmuck, N. Schmidt, and R. Möller, Spring constant of a tuning-fork sensor for dynamic force microscopy. Beilstein Journal of Nanotechnology **3**, 809-816 (2012).

8. J. Falter, M. Stiefermann, G. Langewisch, P. Schurig, H. Hölscher, H. Fuchs, and A. Schirmeisen, Calibration of quartz tuning fork spring constants for non-contact atomic force microscopy: direct mechanical measurements and simulations. Beilstein Journal of Nanotechnology **5**, 507-516 (2014).

9. O.E. Dagdeviren and U.D. Schwarz, Numerical performance analysis of quartz tuning fork-based force sensors. Measurement Science and Technology **28**, 015102 (2017).

10. O.E. Dagdeviren and U.D. Schwarz, Optimizing qPlus sensor assembly for simultaneous scanning tunneling and noncontact atomic force microscopy operation based on finite element method analysis. Beilstein Journal of Nanotechnology **8**, 657-666 (2017).

11. T.G.-C. Whang Y.J. Chris, *A Study of Small Sized Quartz Tuning Fork Using Finite Element Analysis*. 2008: Third Japan-Taiwan Workshop on Future Frequency Control Devices.

12. P. Malatkar, *Nonlinear Vibrations of Cantilever Beams and Plates*. 2003, Virginia Polytechnic Institute and State University.

13. M. Petyt, *Introduction to Finite Element Vibration Analysis*. Second Edition ed. 2010: Cambridge University Press.

14. J.J. Ruz, V. Pini, O. Malvar, P.M. Kosaka, M. Calleja, and J. Tamayo, Effect of surface stress induced curvature on the eigenfrequencies of microcantilever plates. AIP Advances **8**, 105213 (2018).

15. N.A. Burnham, X. Chen, C.S. Hodges, G.A. Matei, E.J. Thoreson, C.J. Roberts, M.C. Davies, and S.J.B. Tendler, Comparison of calibration methods for atomic-force microscopy cantilevers. Nanotechnology **14**, 1 (2003).

16. H.-J. Butt, B. Cappella, and M. Kappl, Force measurements with the atomic force microscope: Technique, interpretation and applications. Surface Science Reports **59**, 1-152 (2005).

17. S.M. Cook, T.E. Schäffer, K.M. Chynoweth, M. Wigton, R.W. Simmonds, and K.M. Lang, Practical implementation of dynamic methods for measuring atomic force microscope cantilever spring constants. Nanotechnology **17**, 2135 (2006).

18. M.L.B. Palacio and B. Bhushan, Normal and Lateral Force Calibration Techniques for AFM Cantilevers. Critical Reviews in Solid State and Materials Sciences **35**, 73-104 (2010).

19. F.J. Giessibl, High-speed force sensor for force microscopy and profilometry utilizing a quartz tuning fork. Applied Physics Letters **73**, 3956-3958 (1998).

20. B.J. Albers, M. Liebmann, T.C. Schwendemann, M.Z. Baykara, M. Heyde, M. Salmeron, E.I. Altman, and U.D. Schwarz, Combined low-temperature scanning tunneling/atomic force microscope for atomic resolution imaging and site-specific force spectroscopy. Review of Scientific Instruments **79**, 033704 (2008).

21. B.-Y. Chen, M.-K. Yeh, and N.-H. Tai, Accuracy of the Spring Constant of Atomic Force Microscopy Cantilevers by Finite Element Method. Analytical Chemistry **79**, 1333-1338 (2007).

22. B. Anczykowski, D. Krüger, K.L. Babcock, and H. Fuchs, Basic properties of dynamic force spectroscopy with the scanning force microscope in experiment and simulation. Ultramicroscopy **66**, 251-259 (1996).

23. O.E. Dagdeviren, J. Götzen, H. Hölscher, E.I. Altman, and U.D. Schwarz, Robust high-resolution imaging and quantitative force measurement with tuned-oscillator atomic force microscopy. Nanotechnology **27**, 065703 (2016).

24. H. Hölscher, Theory of phase-modulation atomic force microscopy with constant-oscillation amplitude. Journal of Applied Physics **103**, 064317 (2008).

25. H. Hölscher and U.D. Schwarz, Theory of amplitude modulation atomic force microscopy with and without Q-Control. International Journal of Non-Linear Mechanics **42**, 608-625 (2007).

26. O.E. Dagdeviren, C. Zhou, E.I. Altman, and U.D. Schwarz, Quantifying Tip-Sample Interactions in Vacuum Using Cantilever-Based Sensors: An Analysis. Physical Review Applied **9**, 044040 (2018).

27. J. Israelachvili, *Intermolecular and Surface Forces*. 2 ed. 1991, London: Academic Press.

28. U.D. Schwarz, A generalized analytical model for the elastic deformation of an adhesive contact between a sphere and a flat surface. Journal of Colloid and Interface Science **261**, 99-106 (2003).

29. D. Maugis and B. Gauthier-Manuel, JKR-DMT transition in the presence of a liquid meniscus. Journal of Adhesion Science and Technology **8**, 1311-1322 (1994).

30. J.A. Greenwood and K.L. Johnson, An alternative to the Maugis model of adhesion between elastic spheres. Journal of Physics D: Applied Physics **31**, 3279 (1998).

31. U.D. Schwarz, O. Zwörner, P. Köster, and R. Wiesendanger, Quantitative analysis of the frictional properties of solid materials at low loads. I. Carbon compounds. Physical Review B **56**, 6987-6996 (1997).

32. J.E. Sader and S.P. Jarvis, Accurate formulas for interaction force and energy in frequency modulation force spectroscopy. Applied Physics Letters **84**, 1801-1803 (2004).